\documentclass[aps,prb,twocolumn,showpacs,groupedaddress]{revtex4-1}
\usepackage{amssymb}

\usepackage{graphicx}

\begin{document}

\title[]{Weak ferromagnetism and internal magnetoelectric effect in LiFeP$_2$O$_7$}

\author{K.-C. Liang$^1$, W. Zhang$^2$, B. Lorenz$^1$, Y. Y. Sun$^1$, P. S. Halasyamani$^2$, and C. W. Chu$^{1,3}$}

\affiliation{$^1$ TCSUH and Department of Physics, University of Houston, Houston, Texas 77204-5002, USA}

\affiliation{$^2$ TCSUH and Department of Chemistry, University of Houston, Houston, Texas 77204-5002, USA}

\affiliation{$^3$ Lawrence Berkeley National Laboratory, 1 Cyclotron Road, Berkeley, California 94720, USA}

\begin{abstract}
The magnetic, thermodynamic, and pyroelectric properties of LiFeP$_2$O$_7$ single crystals are investigated with emphasis on the magnetoelectric interaction of the electrical polarization with the magnetic order parameter. The magnetic order below T$_N\simeq$ 27 K is found to be a canted antiferromagnet with a weak ferromagnetic component along the $b-$axis. A sharp peak of the pyroelectric current at T$_N$ proves the strong internal magnetoelectric interaction resulting in a sizable polarization decrease at the onset of magnetic order. The magnetoelectric effect in external magnetic fields combines a linear and a quadratic field dependence below T$_N$. Thermal expansion data show a large uniaxial magnetoelastic response and prove the existence of strong spin lattice coupling. LiFeP$_2$O$_7$ is a polar compound with a strong interaction of the magnetic order parameter with the electric polarization and the lattice.
\end{abstract}


\maketitle
\section{Introduction}
The coexistence and mutual interaction of magnetic and electric orders in matter has long inspired scientists because of a wealth of novel physical phenomena observed in those systems over the course of more than a century. The magnetoelectric effect, describing the interaction and cross correlation of magnetic (electric) orders with electric (magnetic) fields, has now been revealed in a large number of materials, starting with Cr$_2$O$_3$ in 1959 to 1961.\cite{dzyaloshinskii:59,astrov:60,astrov:61,folen:61} The magnetic and electric fields can be external as well as internal (due to spin or charge orders), the coupling can be linear as well as higher order. A recent review lists a large number of linear magnetoelectric compounds that have been discovered since then.\cite{schmid:03} The internal magnetoelectric coupling is realized in materials with coexisting magnetic and ferroelectric (FE) orders, so called multiferroics.\cite{fiebig:05,spaldin:05,tokura:07,cheong:07}

Multiferroic materials can be separated into two classes. The first class includes compounds in which the FE order is improper and the polarization is induced by an inversion symmetry breaking magnetic order, for example the transverse spin spiral in TbMnO$_3$,\cite{kimura:03} Ni$_3$V$_2$O$_8$,\cite{lawes:05} and MnWO$_4$.\cite{arkenbout:06} The second class includes materials that are already polar (FE) before magnetic order sets in at lower temperature. The classical examples are the hexagonal manganites which become ferroelectric above room temperature and develop a frustrated antiferromagnetic order below 100 K.\cite{fiebig:00,yen:07} A significant coupling of magnetic and dielectric properties has been revealed in those compounds.\cite{lorenz:04,hur:09} In general, any polar material containing magnetic ions with strong interactions leading to magnetic long range order at lower temperatures is a possible candidate that warrants further studies with respect to the the cross coupling of the magnetic order parameter and the electrical polarization.

LiFeP$_2$O$_7$ crystallizes in a polar structure with the monoclinic space group P2$_1$ (No 4). The magnetic ion Fe$^{3+}$ gives rise to antiferromagnetic (AFM) order at about 22 ($\pm$5) K, as shown in earlier studies of polycrystalline samples and powders.\cite{riou:90,rousse:02} The coexistence of the magnetic order with the electrical polarization below the magnetic ordering temperature makes LiFeP$_2$O$_7$ an interesting candidate for studying the magnetoelectric coupling in electrically polarized materials, similar to the multiferrois of the second class, as defined above. The possible interaction of the magnetic order with the electrical polarization in LiFeP$_2$O$_7$ and the effect of external magnetic fields (magnetoelectric effect) are of particular interest and will be investigated in this study.

\section{Synthesis and Experimental}
Large single crystals (centimeter size) of LiFeP$_2$O$_7$ have been grown by a top-seeded solution growth method using a mixture of polycrystalline LiFeP$_2$O$_7$, LiH$_2$PO$_4$, and NH$_4$H$_2$PO$_4$. Details of the crystal growth are described elsewhere.\cite{zhang:12} The phase purity of the single crystals was confirmed by powder X-ray diffraction.

The magnetic properties were measured in a commercial magnetic property measurement system (MPMS, Quantum Design) along different crystallographic orientations. The effect of the magnetic order and external fields on the electrical polarization was determined from integrating the pyroelectric and magnetoelectric currents measured upon changing temperature and field, respectively. The lattice response to the magnetic phase transition was studied via thermal expansion measurements employing a high-resolution capacitance dilatometer. The heat capacity was measured using the relaxation method implemented in the physical property measurement system (PPMS, Quantum Design).

Fig. 1 shows the structure as viewed along the $a-$axis. The almost regular FeO$_6$ octahedra and the diphosphate groups P$_2$O$_7$ form alternating layers stacked along the $b-$axis. The diphosphate groups and the FeO$_6$ octahedra share corners only. The Li ions are located in an off center position in tunnels running along the $a-$axis (Fig. 1). The only magnetic exchange interaction between two neighboring Fe$^{3+}$ ions is the super-super exchange via two oxygen ions involving the stretched PO$_4$ tetrahedron. The relevant magnetic exchange pathways have been discussed recently in more detail.\cite{whangbo:04} The magnetic order of the Fe$^{3+}$ spins (S=5/2) was described based on powder neutron scattering as AFM with the spin orientation mainly along the $a-$axis, although a small component along $c$ provided a better fit to the neutron spectra.\cite{rousse:02} The magnetic modulation vector, $\overrightarrow{q}$=(0,0,0), only allows for the AFM spin order of the two magnetic ions within the unit cell, the magnetic unit cell is therefore identical to the crystallographic cell.

\begin{figure}
\begin{center}
\includegraphics[angle=0, width=3 in]{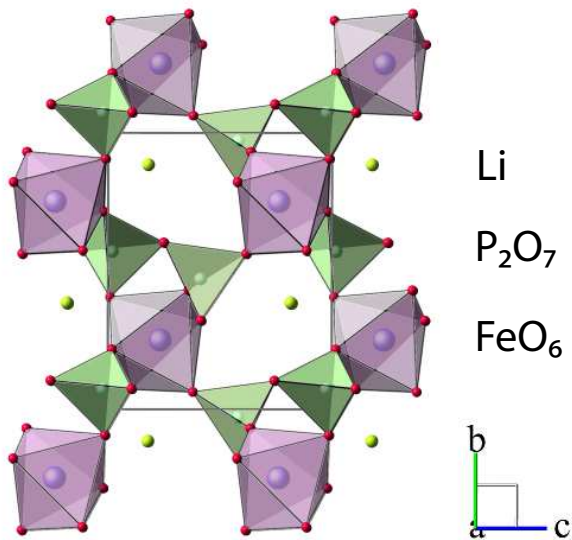}
\end{center}
\caption{(Color online) Structure of a LiFeP$_2$O$_7$ viewed along the $a$-axis.}
\end{figure}

\section{Results and discussion}
\subsection{Magnetic order below T$_N$=27 K}
The single crystals of LiFeP$_2$O$_7$ grown for this study permit a more detailed investigation of the magnetic properties. In contrast to the earlier reports,\cite{riou:90,rousse:02} the magnetic moment M$_b$ measured in low fields (10 Oe) along the $b-$axis shows a sharp increase at T$_N\simeq27$ K, indicating a ferromagnetic (FM) moment along this direction (Fig. 2a). The data taken after cooling in zero magnetic field (ZFC) and cooling in a 10 Oe field (FC) used for the measurement of the magnetization are identical. This appears unusual since ZFC and FC data of a ferromagnet are commonly different because of magnetic domain formation. However, if the magnetically ordered state is a single domain state, both data sets will be identical and the ferromagnetic M-H hysteresis loop should be very sharp. The inset of Fig. 2a shows the ac-susceptibility, $\chi_b'$ (measured at 117 Hz with a field amplitude of 1 Oe), in a narrow temperature range near T$_N$, between 26.7 K and 27.1 K. The sharp peak of $\chi_b'$ with a width of only 0.01 K is consistent with the second order transition at the critical temperature T$_N$=26.86 K.

The single domain state is further supported by M-H measurements shown at different temperatures in Fig. 2b. The FM hysteresis loops are indeed very sharp, even at temperatures as high as 25 K, just below $T_N$. The reversal of the FM moment at the coercive field happens instantaneously within the data resolution. Some data have been collected in field steps as low as 10 Oe and no intermediate magnetization values could be detected near the coercive field, i.e. the spontaneous FM moment flips at once from positive to negative values and vice versa, as indicated by the vertical dashed lines in Fig. 2b. Once the direction of the FM moment is set, the $b-$axis magnetization further increases linearly with the field up to the maximum field of 50 kOe. The magnetic hysteresis loops show a small asymmetry the origin of which is not clear. It could be related to the complex interaction of the FM order parameter $M^Y$ with the AFM components $L^X$, $L^Z$, and the polarization $P^Y$, as discussed in Section III D below.

\begin{figure}
\begin{center}
\includegraphics[angle=0, width=3 in]{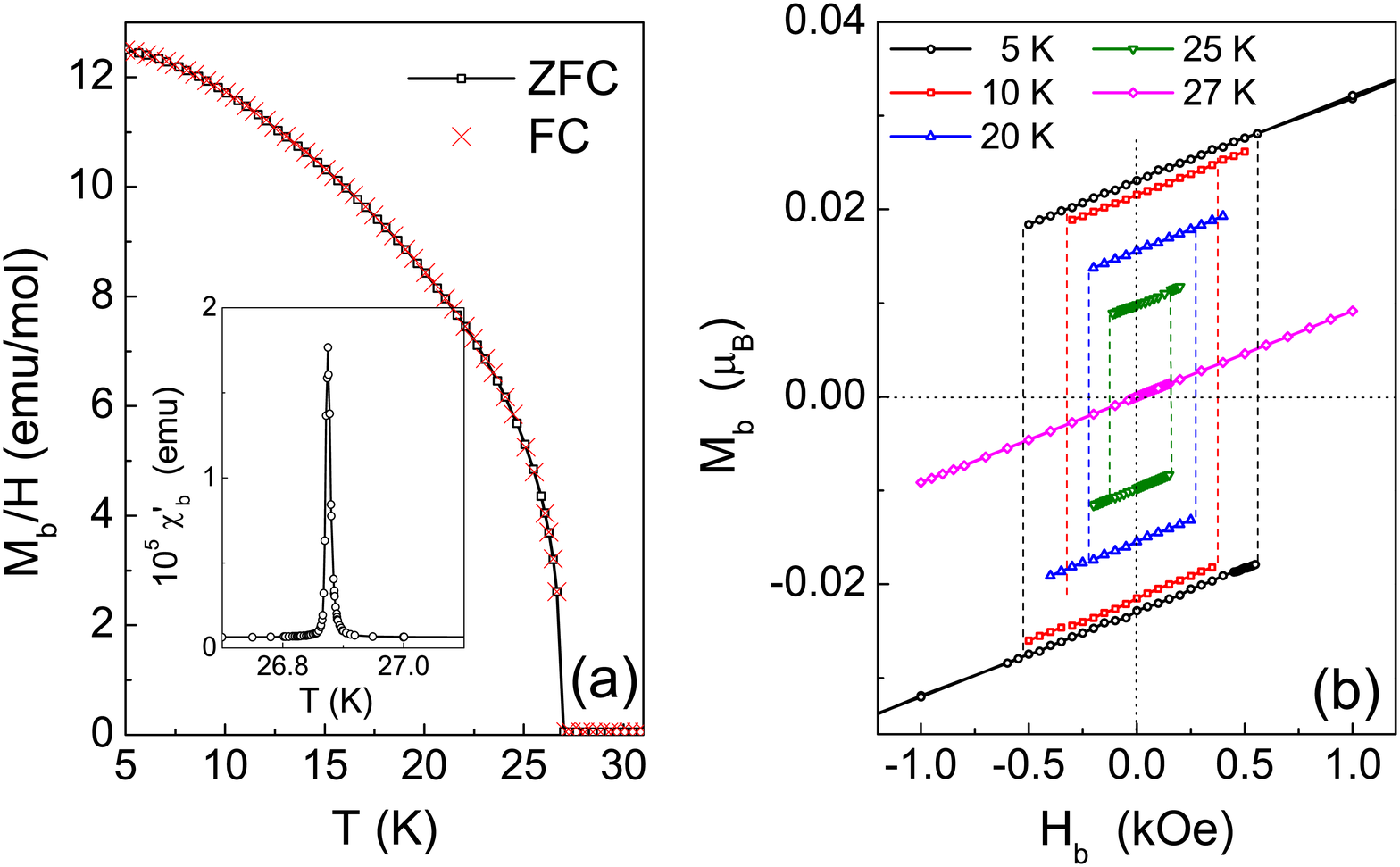}
\end{center}
\caption{(Color online) $b-$axis magnetization of LiFeP$_2$O$_7$. (a) M$_b$ vs. T at 10 Oe. (b) M$_b$ vs. H at different temperatures. Note that the zero-field cooled (ZFC) and field cooled (FC) data in (a) perfectly overlap indicating the single domain property of the magnetic order. The inset in (a) shows the ac susceptibility near the phase transition.}
\end{figure}

The magnetic properties measured along the $b-$axis suggest that the true magnetic state of LiFeP$_2$O$_7$ below T$_N$ is a canted antiferromagnet with the spins tilted slightly from the $a-c$ plane towards the $b-$axis, requiring a revision of the earlier results obtained from powder samples\cite{riou:90,rousse:02} and theoretical considerations.\cite{whangbo:04} This canted spin picture is confirmed by comparing the magnetization measured along different crystallographic orientations. Fig. 3 displays the magnetization data at 1000 Oe along the $a-$, $b-$ and $c-$axes. M$_b$ shows the FM response similar to the low-field data, however, M$_a$ decreases sharply at T$_N$, as expected for an antiferromagnet. M$_c$ experiences only a very minute anomaly at T$_N$ and remains nearly constant at lower temperatures. The Curie-Weiss extrapolation of the high-temperature data reveals a negative Weiss temperature $\Theta\simeq-$50 K and an effective magnetic moment of $\mu_{eff}\simeq$ 6 $\mu_B$, close to the expected value for spin 5/2. The negative Weiss temperature proves that the dominant magnetic interactions are antiferromagnetic in nature and the FM moment along the $b-$axis most likely originates from the canted AFM order.

It should be noted that the FM component of the magnetic order parameter does not violate any symmetry constraints. According to the neutron scattering data of LiFeP$_2$O$_7$ powders, the magnetic modulation vector is $\overrightarrow{q}$=(0,0,0) and the magnetic group of $\overrightarrow{q}$ can be decomposed into two irreducible representations, $\Gamma$=3($\Gamma_1$+$\Gamma_2$).\cite{rousse:02} The basis vectors of the $\Gamma_1$ representation include two AFM components ($L^X=S_1^X-S_2^X$ and $L^Z=S_1^Z-S_2^Z$), both observed in neutron scattering data, and one FM component ($M^Y=S_1^Y+S_2^Y$). $\overrightarrow{S}_1$ and $\overrightarrow{S}_2$ refer to the spin vectors of the two iron ions in the unit cell. $M^Y$ allows for the ferromagnetic moment along the $b-$axis, as observed in our measurements. The fact that the FM rise of M$_b$ and the AFM drop of M$_a$ happen at exactly the same temperature T$_N$ proves that both are components of one and the same magnetic order parameter. There remains the question why the $b-$axis FM moment has not been detected in earlier magnetization\cite{riou:90} and neutron scattering\cite{rousse:02} experiments. In both previous studies, polycrystalline or powder samples had been investigated. Magnetic measurements of randomly oriented particles can only detect a magnetization averaged over all orientations and the FM moment could have escaped the attention. Neutron scattering measurements have a finite resolution limit to detect the size of magnetic moments. The spontaneous FM moment measured at 5 K is as small as 0.024 $\mu_B$/Fe (Fig. 2b). Therefore, powder neutron scattering experiments could not detect the FM component of the order parameter.

\begin{figure}
\begin{center}
\includegraphics[angle=0, width=3 in]{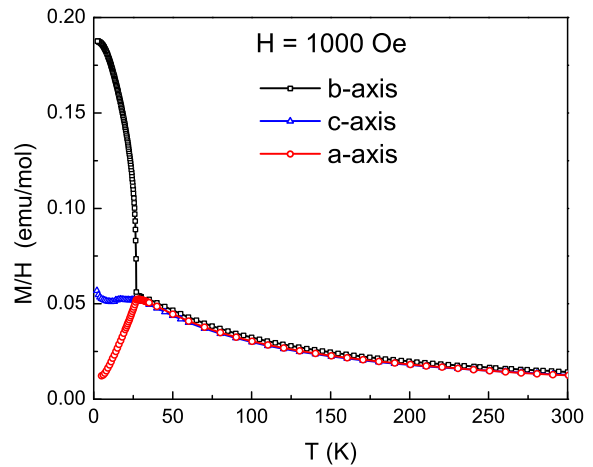}
\end{center}
\caption{(Color online) Magnetization M/H of LiFeP$_2$O$_7$ measured at 1000 Oe along the $a-$, $b-$ and $c-$axes.}
\end{figure}

The single domain property and the smooth increase of the FM order parameter at and below T$_N$ strongly suggest the second order character of the magnetic phase transition. This is further supported by the results of heat capacity (C$_p$) measurements, shown in Fig. 4. C$_p$(T) exhibits a very sharp peak at T$_N$, similar to the $\lambda$-shaped anomaly of C$_p$ that is frequently observed near second order phase transitions if critical fluctuations of the order parameter are significant. Since the magnetic phase transition is continuous, the expansion of the free energy with respect to the order parameters will provide a mean field description of the magnetic phase transition, as discussed in see Section III D.

\begin{figure}
\begin{center}
\includegraphics[angle=0, width=3 in]{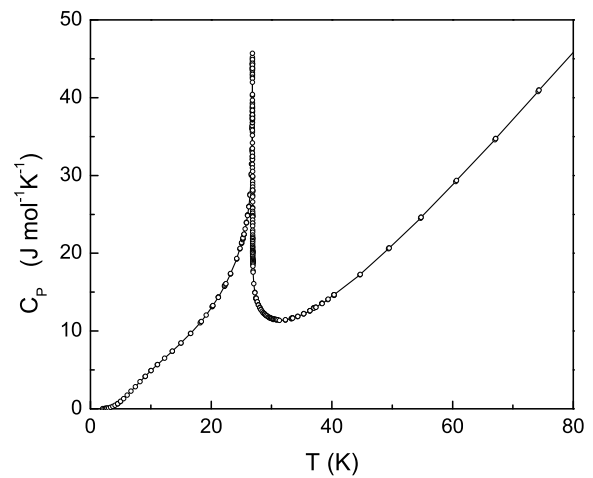}
\end{center}
\caption{Heat capacity of LiFeP$_2$O$_7$ near the magnetic phase transition.}
\end{figure}

\subsection{Pyroelectric and magnetoelectric response of LiFeP$_2$O$_7$}
Because of the polar structure of LiFeP$_2$O$_7$, a macroscopic electrical polarization exists at all temperatures below ambient. However, it is not clear whether or not the polar distortion disappears at some higher critical temperature resulting possibly in a structural and ferroelectric transition. Polarization measurements at ambient temperature in electric fields up to 24 kV/cm did not detect any hysteresis in the P(E) data.\cite{zhang:12} High temperature data for LiFeP$_2$O$_7$ are rare. Magnetic data collected up to 700 K have not indicated any additional anomaly that could be associated with a high-temperature phase transition.\cite{riou:90} Although high-temperature measurements of the ionic conductivity of LiFeP$_2$O$_7$ have shown a crossover near 700 K, differential thermal analysis data have not shown any notable anomaly that might suggest a structural or ferroelectric transition.\cite{vitins:00}

The single crystals used in this study have been found to be good insulators at and below ambient temperatures. Therefore, the change of the electrical polarization can be measured using the pyroelectric current method. Current data acquired upon increasing and decreasing temperature (rate of temperature change: 3 K/min) along the axis of polarization ($b-$axis) are identical in magnitude but with opposite sign, proving that the detected current is in fact due to the intrinsic polarization change. Fig. 5 shows the pyroelectric response upon increasing temperature. The measured current is significant and it proves the strong increase of the polarization towards lower temperatures. The polarization change between ambient and low temperatures, $\Delta$P$_b$(T), is obtained by integrating the current and dividing by the contact area and it is shown in the inset of Fig. 5.

The most prominent feature of the pyroelectric current I$_p$(T) is its sharp peak at T$_N\simeq$ 27 K, the critical temperature of the magnetic phase transition, resulting in the drop of the polarization below T$_N$. This pronounced peak implies a very strong coupling of the magnetic order parameter and the $b-$axis polarization, i.e. a large internal magnetoelectric effect. A closer inspection of the pyroelectric data also shows that $\Delta$P$_b$(T) reaches its maximum (I$_p$=0) at much higher temperatures of about 2 T$_N$, which can be understood as the onset of sizable magnetic fluctuations and their effect on the lattice. This is also consistent with the heat capacity data showing an enhancement of C$_p$ over the lattice contribution, starting  at about 50 K (Fig. 4).

The external magnetoelectric effect (the change of P$_b$ in applied magnetic fields) was studied at different temperatures and the results are shown in Fig. 6. The $b-$axis polarization increases in longitudinal fields H$_b$ below the magnetic transition temperature as well as above (35 K data in Fig. 6a). However, the magnitude of $\Delta$P$_b$(H) is small at low fields but it increases nonlinearly at higher magnetic field. It is interesting that $\Delta$P$_b$(H) appears to be linear at low fields for temperatures below T$_N$. This is shown more clearly by plotting the magnetoelectric current I$_{me}$ (measured at 200 Oe/s) in Fig. 6b. I$_{me}$ increases linearly with H$_b$ at all temperatures, however, the intercept with the vertical axis is finite only at temperatures below T$_N$. The dashed lines in Fig. 6b are a linear fit to the data and they show the finite intercept for the 5 K and 15 K data, but not for the 35 K data. Below T$_N$, I$_{me}$ jumps to a finite value as soon as the field starts to increase. This is a clear indication of a linear magnetoelectric effect which is superimposed on the quadratic magnetoelectric coupling that exists at all temperatures.

The superposition of linear and quadratic magnetoelectric effects below T$_N$ is a consequence of the spontaneous magnetic order which couples linearly to the field and to the polarization, as discussed in more detail in Section III D below. It should be noted that the polarization change is symmetric with respect to the field orientation and data measured at 5 K in negative magnetic fields have been included as stars in Fig. 6a. Similar measurements in a transverse magnetic field have not shown any significant change of the $b-$axis polarization. The corresponding magnetoelectric current was an order of magnitude smaller and close to the resolution limit of the measurement.

\begin{figure}
\begin{center}
\includegraphics[angle=0, width=3 in]{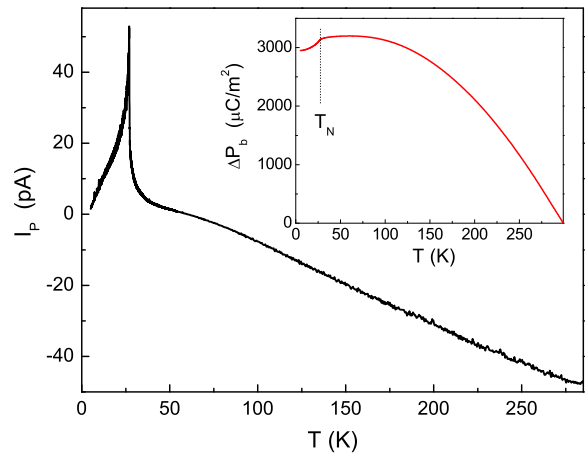}
\end{center}
\caption{(Color online) Pyroelectric current and polarization change (inset) vs. temperature of LiFeP$_2$O$_7$.}
\end{figure}

\begin{figure}
\begin{center}
\includegraphics[angle=0, width=3 in]{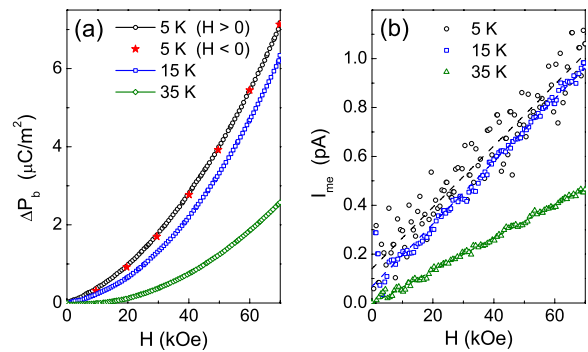}
\end{center}
\caption{(Color online) (a) Magnetoelectric polarization change, $\Delta P_b(H)$, and (b) magnetoelectric current, $I_{me}$, at three different temperatures. The bold stars in (a) show the polarization change in a negative magnetic field. The dashed lines in (b) are a linear fit of the magnetoelectric current data.}
\end{figure}

\subsection{Lattice anomalies at the magnetic transition}
The strong effect of the magnetic order below T$_N$ on the lattice polarization requires the presence of significant spin-lattice coupling resulting in ionic displacements with the corresponding change of the electrical polarization, as shown in Fig. 5. The macroscopic distortion of the lattice below T$_N$ is revealed through thermal expansion measurements. The relative length changes $\Delta L(T)/L_0$ of three orthogonal directions, chosen as the $b-$ and $c-$axes of the crystal and the orientation perpendicular to both axes, are shown in Fig. 7 at low temperatures. It is remarkable that the crystal's dimensions along $b-$ and $\perp (b,c)-$ directions change very little in passing through the magnetic phase transition. However, the crystal exhibits a significant contraction of the $c-$axis with a pronounced anomaly at T$_N$. The thermal expansivity of the $c-$axis, $\alpha=\partial (\Delta c/c_0)/\partial T$, increases sharply near T$_N$ and displays a $\lambda$ shaped peak, similar to the pyroelectric current and the heat capacity. This demonstrates the strong interaction of the magnetic order with the lattice.

It is not clear at this point why the lattice response to the magnetic order is nearly uniaxial along the $c-$direction only (Fig. 7). From the structure of LiFeP$_2$O$_7$ it is obvious that the diphosphate group is partially oriented along the direction of contraction. It appears therefore conceivable to assume that the magnetic order mainly distorts the diphosphate group to maximize the energy gain of the magnetic system. A decrease of the angle P-O-P within the diphosphate group from 129$^\circ$ will also result in a contraction of the $c-$axis. The distortion of the PO$_4$ tetrahedra results in a change of the shape of the tunnels running along the $a-$direction. Therefore, a displacement of the Li ions in the tunnels can be expected resulting in the observed decrease of the polarization below T$_N$.

\begin{figure}
\begin{center}
\includegraphics[angle=0, width=3 in]{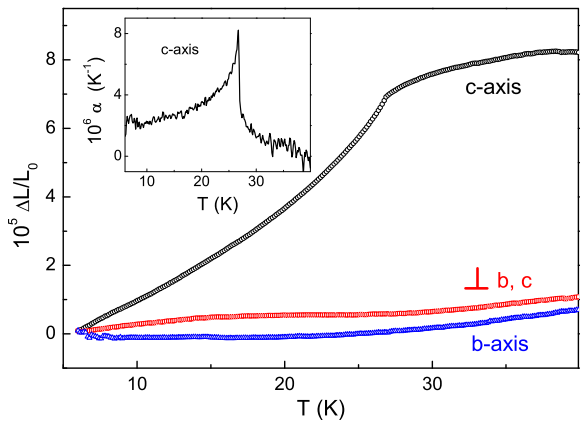}
\end{center}
\caption{(Color online) Relative length change of $b$, $c$, and perpendicular ($\bot$) to $b$ and $c$ with reference to 6 K. The inset shows the $c-$axis thermal expansivity $\alpha$(T).}
\end{figure}

The distortion of the PO$_4$ units below the magnetic phase transition will change the super-super exchange interactions which depend strongly on the O-O distance along the exchange path Fe-O-O-Fe.\cite{whangbo:04} The observed lattice contraction (Fig. 7) is therefore a result of a competition between the gain of magnetic exchange energy due to a distortion-induced modification of the super-super exchange parameters and the elastic energy of the crystal. Theoretical estimates of the exchange coupling parameters\cite{whangbo:04} suggest that the strongest coupling ($J_3$) is found between the two Fe spins $\overrightarrow{S}_1$ and $\overrightarrow{S}_2$ in the unit cell and a sizable coupling ($J_4$) exists between $\overrightarrow{S}_1$ and $\overrightarrow{S}_2$ of two neighboring unit cells stacked along the $a-$axis (here we use the same labels for the $J_i$ as in the calculation of Ref. \cite{whangbo:04}). The corresponding exchange parameters are schematically shown in Fig. 8. The contraction along the $c-$axis will reduce the O-O distance of these two super-super exchange pathways, increase the magnetic exchange constants, and maximize the energy gain in the ordered state. While this discussion based on the macroscopic expansion anomalies can only be qualitative, high-resolution X-ray or neutron scattering experiments could resolve the microscopic details of the lattice distortion and lead to a more fundamental understanding of the observed effects.

\subsection{Landau free energy expansion}
The second order character of the magnetic phase transition is well established through the magnetic (Fig. 2a) and thermodynamic (Fig. 4) data. Within a mean field description, the Landau free energy can therefore be expanded with respect to the magnetic order parameter. The possible coupling of the components of the magnetic order parameter and the lattice polarization will be included. All terms in the free energy expansion have to be invariant with respect to the symmetry operations of the crystal above T$_N$ and the time reversal operation. The crystal's space group P2$_1$ has only two symmetry elements, the identity and the screw type operation (180$^\circ$ rotation about the $b-$axis and a translation by $\overrightarrow{b}$/2). Based on previous and current data, the magnetic order parameter in the little group of $\overrightarrow{q}$=(0,0,0) is defined by the $\Gamma_1$ irreducible representation as ($L^X$, $L^Z$, $M^Y$). The free energy expansion in terms of the magnetic order parameter and the electrical polarization is given by:

\begin{widetext}
\begin{eqnarray}
F(\overrightarrow{L},\overrightarrow{M},\overrightarrow{P}) = F_0+{a\over 2}L^2+{b\over 4}L^4+{c\over 2}M^2+{d\over 4}M^4+\sigma_1L^XL^Z+\sigma_2L^XM^Y+\sigma_3L^ZM^Y-\alpha P^Y+{\beta\over 2}(P^Y)^2\nonumber\\
+[\lambda_1L^XL^Z+\lambda_2L^XM^Y+\lambda_3L^ZM^Y+\lambda_4(L^X)^2+\lambda_5(L^Z)^2+\lambda_6(M^Y)^2]P^Y+...
\end{eqnarray}
\end{widetext}

\begin{figure}
\begin{center}
\includegraphics[angle=0, width=2.5 in]{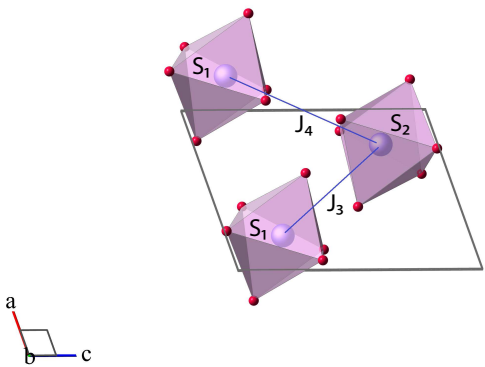}
\end{center}
\caption{(Color online) Schematic plot of the most relevant super-super exchange interactions, J$_3$ and J$_4$, between different iron moments (only FeO$_6$ octahedra are shown).\cite{whangbo:04}  }
\end{figure}

Equation (1) includes the second and fourth orders of the magnetic order parameters $\overrightarrow{L}$, $\overrightarrow{M}$ and all other terms up to third order that are invariant under the symmetry operations of the crystal and the time reversal. The first line represents the magnetic system, limited to the components of the order parameter in $\Gamma_1$ representation, and the energy of the polarized state. Note that the linear term in $P^Y$ is symmetry allowed in the P2$_1$ polar structure. The $\sigma_1$, $\sigma_2$, and $\sigma_3$ terms couple different components of the magnetic order parameter and they are commonly of relativistic origin. Neutron scattering experiments\cite{rousse:02} and our magnetization data (Figs. 2, 3) show that the major magnetization component in the ordered state is $L^X$ whereas $L^Z$ and $M^Y$ are comparatively small. This leads us to conclude that the primary order parameter is $L^X$ and the two other components are coupled to $L^X$ through the bilinear terms $\sigma_1$ and $\sigma_2$, resulting in the increase of all three components of the magnetic order parameter at the same critical temperature, T$_N$. This conclusion is further supported by the magnetization data of Fig. 3. The $a-$axis magnetization shows a clear antiferromagnetic response below T$_N$ with a drastic decrease to lower temperatures. However, the $c-$axis response is far more subtle and a significant decrease below T$_N$ is missing, indicating that the $L^Z$ component of the order parameter is most probably induced by the weak coupling through the $\lambda_1L^XL^Z$ term in equation (1). The second line in (1) shows the symmetry allowed third order terms which couple the magnetic order and the polarization.

Minimizing the free energy (1) with respect to $L^X$, $L^Z$, $M^Y$, and $P^Y$ will define the thermodynamically stable state and the magnetic/dielectric orders as functions of temperature. However, because of the large number of unknown parameters like a, b,..., $\sigma_i$, and $\lambda_i$, we will not attempt to fit a magnetic solution to the experimental data. The electrical polarization can be derived as:

\begin{eqnarray}
P^Y={1\over \beta}[\alpha-\lambda_1L^XL^Z-\lambda_2L^XM^Y-\lambda_3L^ZM^Y\nonumber\\
-\lambda_4(L^X)^2-\lambda_5(L^Z)^2-\lambda_6(M^Y)^2]
\end{eqnarray}

Equation (2) describes the dependence of the polarization on the magnetic order parameter. The first term, $\alpha/\beta$, represents the temperature dependent polarization in the paramagnetic state, as obtained by integrating the pyroelectric current above T$_N$ (inset, Fig. 5). The remaining terms in (2) describe the response of the electrical polarization to the magnetic phase transition resulting in the peak anomaly of the pyroelectric current and the decrease of the polarization P$_b$ below T$_N$. Note that the significant change of the $c-$axis length below T$_N$, as shown in Fig. 7, is not considered in equations (1) and (2). A more detailed theory that includes the magnetoelastic effects on a microscopic level has yet to be developed.

To include the symmetry allowed terms that couple the magnetic order parameter to the lattice strain tensor $\varepsilon_{ij}$, we have to build invariants of the lowest order in $L^X$, $L^Z$, $M^Y$, and $\varepsilon_{ij}$. Since the lowest order terms have to be bilinear in the components of the magnetic order parameter to preserve the time reversal symmetry, they include all combinations of $L^X$, $L^Z$, and $M^Y$ shown in the square brackets of equation (1). Al those terms are also invariant with respect to the two symmetry elements of the space group, the identity and the screw operation. Therefore, only space-group invariant components $\varepsilon_{ij}$ of the strain tensor are allowed to couple in first order to the bilinear products of the magnetic order parameter components. Those elements are $\varepsilon_{xx}$, $\varepsilon_{yy}$, $\varepsilon_{zz}$, and $\varepsilon_{xz}$. The thermal expansion data shown in Fig. 7 indicate that the major lattice distortion is a compressive strain along the $c$-axis, ruling out any significant contributions from $\varepsilon_{xz}$ (shear strain) and $\varepsilon_{yy}$ (compression of the $b$-axis). The important terms are $\varepsilon_{xx}$ and $\varepsilon_{zz}$ which can account for the observed $c$-axis compression below T$_N$.

The magnetoelectric effects shown in Fig. 6 deserve a more detailed discussion. The coupling of the magnetic order parameter, the magnetic field, and the polarization is of interest. Any product of $L^X$, $L^Z$, and $M^Y$ with the external field $H^Y$ is invariant under the operations of the space group and the time reversal. The polarization $P^Y$ is also an invariant. Therefore, the magnetoelectric effect is described by trilinear terms in the free energy expansion, for example $E_{me}\propto M^Y\cdot H^Y\cdot P^Y$. In the high-temperature phase (T$>$T$_N$), $M^Y=\chi^Y\cdot H^Y$ ($\chi^Y=$ d$M^Y/$d$H^Y$ is the magnetic susceptibility), and the magnetoelectric response is quadratic in the external field,

\begin{eqnarray}
E_{me}\propto\chi^Y(H^Y)^2P^Y
\end{eqnarray}

This is indeed observed above T$_N$, see for example the 35 K data in Fig. 6.

Below T$_N$, however, the existence of a spontaneous magnetization $M^S$ adds an additional term to equation (3). With $M^Y=M^S+\chi^Y\cdot H^Y$,

\begin{eqnarray}
E_{me}\propto M^SH^YP^Y+\chi^Y(H^Y)^2P^Y
\end{eqnarray}

The existence of a spontaneous magnetization below T$_N$ now gives rise to a linear magnetoelectric response in addition to the quadratic effect, in perfect agreement with the experimental results presented in Fig. 6.

\begin{figure}
\begin{center}
\includegraphics[angle=0, width=3 in]{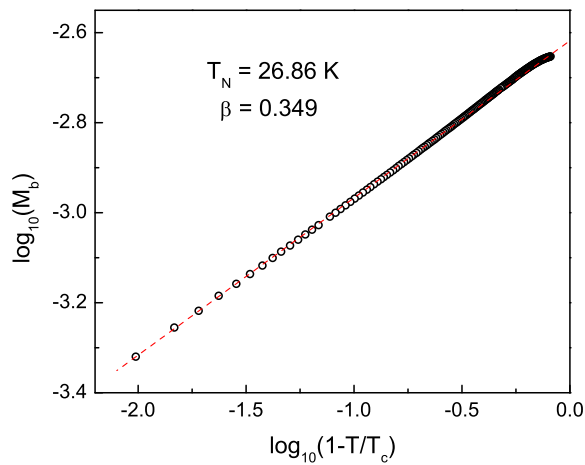}
\end{center}
\caption{(Color online) Scaling plot of the magnetization of LiFeP$_2$O$_7$. The critical exponent $\beta$ is determined by the slope (dashed line).}
\end{figure}

\subsection{Critical scaling of the magnetization}
The mean field theory cannot explain the $\lambda$ type anomaly of the heat capacity since the critical exponent of C$_p$ in the Landau mean field theory is zero. Spatial fluctuations of the order parameter which become critical near T$_N$ are essential. Evidence for critical behavior is derived qualitatively from the $\lambda$-shaped anomaly of the heat capacity and more quantitatively from the temperature dependence of the magnetization (Fig. 2a). M(T) is expected to follow the relation $M(T)=M_0\cdot(1-T/T_N)^\beta$ near the critical temperature. $\beta$=1/2 is the critical exponent of the order parameter in a mean field description; however, a more realistic model description taking into account the spatial fluctuations of the order parameter leads to different values of $\beta$ depending on the model, the spatial dimension, and the specifics of the order parameter.\cite{ma:76} The double logarithmic plot in Fig. 9 shows a significant range where the scaling formula for M(T) is fulfilled, as indicated by the dashed line. The estimated critical temperature T$_N$=26.86 K agrees well with the T$_N$ determined from the sharp ac susceptibility peak. The critical exponent $\beta$=0.349 is clearly smaller than the mean field value but it is consistent with values obtained for typical spin models (Ising or Heisenberg) when critical fluctuations are allowed.\cite{ma:76} This shows the limits of the mean field theory in describing the second order magnetic phase transition. A more advanced description has to take into account the spatial fluctuations of the order parameter and the critical behavior as well as the details of the magnetic order ($L^X$, $L^Z$, $M^Y$) and the microscopic interactions.

\section{Summary}
Single crystals of LiFeP$_2$O$_7$ have been investigated with respect to the magnetic phase transition at T$_N\simeq$ 27 K and their polarization properties below room temperature. We show that the magnetic structure below T$_N$ is more complex than previously discussed. The Fe spins exhibit a canted antiferromagnetic arrangement with a symmetry allowed ferromagnetic component along the crystallographic $b-$axis. Pyroelectric measurements reveal the electrical polarization increasing between room temperature and T$_N$. A significant internal magnetoelectric effect is found in form of a sharp $\lambda$ shaped peak anomaly of the pyroelectric current at the magnetic transition temperature. A similar anomaly of the heat capacity and the continuous increase of the ferromagnetic component of the magnetic order parameter prove the second order nature of the phase transition. The lattice response to the magnetic transition is uniaxial with a clear anomaly of the thermal expansivity measured along the $c-$axis. All experimental data suggest that LiFeP$_2$O$_7$ is a polar crystal with a strong interaction of the magnetic order parameter and the electrical polarization. The magnetoelectric effect is shown to exist in second order of the field above and below the Ne\'{e}l temperature with an additional linear component in the magnetically ordered state. The trilinear magnetoelectric coupling terms contributing to the Landau free energy expansion are derived from symmetry considerations.

\begin{acknowledgments}
We wish to thank Clarina R. dela Cruz for valuable discussions. This work is supported in part by the US Air Force Office of Scientific Research, the T.L.L. Temple Foundation, the J. J. and R. Moores Endowment, and the
State of Texas through the Texas Center for Superconductivity at the University of Houston and at LBNL by the US Department of Energy. W. Z. and P. S. H. thank the Welch Foundation (Grant E-1457) and the Texas Center for Superconductivity for support.
\end{acknowledgments}

%


\end{document}